# IMPLICATIONS OF THE STRANGE SPIN OF THE NUCLEON FOR THE NEUTRON ELECTRIC DIPOLE MOMENT IN SUPERSYMMETRIC THEORIES

**John Ellis**

CERN, CH-1211 Geneva 23, Switzerland

and

**Ricardo A. Flores**

Department of Physics and Astronomy,
University of Missouri,
St. Louis, MO 63121, USA

## ABSTRACT

Supersymmetric model contributions to the neutron electric dipole moment arise via quark electric dipole moment operators, whose matrix elements are usually calculated using the Naive Quark Model (NQM). However, experiments indicate that the NQM does not describe well the quark contributions $\Delta q$ to the nucleon spin, and so may provide misleading estimates of electric dipole operator matrix elements. Taking the $\Delta q$ from experiment, we indeed find consistently smaller estimates of the neutron electric dipole moment for given values of the supersymmetric model parameters. This weakens previous constraints on CP violation in supersymmetric models, which we exemplify analytically in the case where the lightest supersymmetric particle (LSP) is a $U(1)$ gaugino $\tilde{B}$, and display numerically for other LSP candidates.

The fact that the neutron electric dipole moment $d_n$ is expected to be so small in the Standard Model means that it is an excellent window for possible physics beyond the Standard Model, which may cast valuable light on the still–mysterious phenomenon of CP violation. Of special interest to us is the information that $d_n$ may provide on supersymmetry, which has been a focus of attention for quite some time [1], and in particular on the minimal supersymmetric extension of the Standard Model (MSSM). The MSSM contains many possible sources of CP violation beyond the Kobayashi–Maskawa phase of the Standard Model, whose presence could have significant impact on studies of supersymmetric candidates for cold dark matter [2].

Calculations of $d_n$ involve two aspects: a short–distance part that may be calculated perturbatively in the Standard Model or one's chosen extension of it, and a long–distance non–perturbative QCD calculation of the matrix element(s) of the short-distance operator(s). In the case of the MSSM, it is generally thought that the value of $(d_n)$ is dominated by quark electric dipole operators $\frac{1}{4}\bar{q}\sigma_{\mu\nu}q\tilde{F}^{\mu\nu}$ [3], whose short–distance coefficient is related, in particular, to the chargino/neutralino sector that also features in dark–matter calculations.

In the past, the matrix elements of these quark electric dipole moment operators have generally been estimated using the naive quark model (NQM). However, although this model is usually fairly reliable, there are indications that its predictions for the spin content of the nucleon are misleading [4]. In particular, there appears to be a non–zero strange quark contribution $\Delta s$ to the nucleon spin, and the total quark contribution $\Delta u + \Delta d + \Delta s$ is quite small. These observations suggest that the NQM may not be adequate to describe the contributions of the quark electric dipole operators to the neutron electric dipole moment. In particular, a contribution from the strange quark could be significant, and there might be a systematic cancellation between the $u, d$ and $s$ contributions. This would mean that previous constraints on CP–violating parameters in the MSSM derived from upper limits on $d_n$ should be relaxed.

Here we study this possibility analytically in the idealized case where the lightest supersymmetric particle (LSP) is a $U(1)$ gaugino $\tilde{B}$, and numerically over the $(\mu, M_2)$ plane that characterizes the neutralino and chargino masses in the MSSM. We do indeed find a systematic suppression of $d_n$ compared to the NQM prediction for the same values of the CP–violating parameters in the MSSM. This in turn means that the corresponding upper bounds on these parameters inferred from the experimental upper limit on $d_n$ should be relaxed. This provides some breathing space for CP violation in the MSSM that may not be unwelcome, and in turn increases further the upper limit on the LSP mass inferred from calculations of its relic density that allow for CP–violating phases [2].‡

We first discuss the contribution of quarks to $d_n$ in terms of their contribution to the total neutron spin. The dipole operator $\frac{1}{4}\bar{q}\sigma_{\mu\nu}q\tilde{F}^{\mu\nu}$ yields a contribution $d_q(\Delta q)_n$ to $d_n$, where $\langle n|\frac{1}{2}\bar{q}\gamma_\mu\gamma_5 q|n\rangle \equiv (\Delta q)_n S_\mu$ defines the fractional contribution of the quark to the neutron spin $S_\mu$, and $d_q$ is the fundamental quark electric dipole moment that arises due to CP violation. The fractions $(\Delta q)_n$ are related by isospin symmetry to the fractions $\Delta q$ of the proton: $(\Delta u)_n = \Delta d$, $(\Delta d)_n = \Delta u$, and $(\Delta s)_n = \Delta s$ for the $u$, $d$ and $s$ quarks. In the non–relativistic NQM one would have $\Delta u = \frac{4}{3}$, $\Delta d = -\frac{1}{3}$, and $\Delta s = 0$. Thus, $d_n = \frac{4}{3}d_d - \frac{1}{3}d_u$ in this case, which is the relation that has always been used in the previous literature on this subject. However, there are good indications that neither this model nor a simple relativistic modification of the NQM are adequate to describe the spin structure of the nucleon. Recent results [4] have confirmed the earlier result by the European Muon Collaboration (EMC) [5] that the total quark contribution $\Delta u + \Delta d + \Delta s$ is quite small. An analysis of the combined deep inelastic scattering and hyperon $\beta$–decay data yields $\Delta u = 0.82 \pm 0.03$, $\Delta u = -0.44 \pm 0.03$, and $\Delta s = -0.11 \pm 0.03$ [4]. We shall refer to these as EMC values of the fractions $\Delta q$.

It is easy to see how the EMC values of the spin fractions tend to give smaller values of $d_n$. Since $d_q \propto m_q$, to a good approximation we have that $d_n \propto m_d \Delta u + m_s \Delta s$. This is so because $|\Delta u| \approx 2|\Delta d|$, and $m_d \approx 2m_u$, therefore the $u$ quark contribution is subdominant. Then, since the $d$ and $s$ quarks have the same quantum numbers, and $m_s \approx 150$ MeV $\gg m_d \approx 10$ MeV, we have contributions of approximately equal magnitude and opposite sign for small and negative values of $\Delta s$, as indicated by the EMC.

---

‡Note that, in general, constraints based on the upper limit to the electron's electric dipole moment may be similar to those based on $d_n$, if one assumes similar squark and slepton masses.

We now turn to the calculation of $d_n$ in the MSSM to show that these expectations are borne out. The calculation of the fundamental quark electric–dipole moments in the MSSM has been extensively discussed in the literature. Here we shall follow the phase conventions and complete one–loop analysis of ref. [6] in order to calculate $d_n$ from the fundamental quark moments $d_q$.

We begin our discussion with the simple case in which the LSP is the pure gaugino $\tilde{B}$. This is a good approximation to the composition of the MSSM LSP in the region $|\mu| \gg M_2$ of the parameter space characterized by the magnitude of the Higgs mixing mass $\mu$ and SU(2) gaugino mass $M_2$ (for simplicity, we shall assume GUT conditions on the gaugino masses so that we need to consider only one of them), and where the LSP makes an attractive dark–matter candidate. The quark contributions, $d_q$, to $d_n = \sum_{q \in n} d_q (\Delta q)_n$ involve only gluino and chargino exchange if we neglect the possible contributions of heavier neutralinos. Note that in the MSSM there is no net contribution to $d_q$ from the heavier neutralinos in the region $|\mu| \gg M_2$, so this is not an unrealistic model. The contribution due to gluino exchange is given by (assuming a common squark mass; we shall relax this assumption in our later numerical analysis)

$$d_q^G/e = \frac{2\alpha_s}{3\pi} Q_q \frac{\overline{m}_q \sin\gamma_q}{M_{\tilde{q}}} \frac{m_q}{M_{\tilde{q}}^2} \sqrt{r} K(r) , \qquad (1)$$

where $\overline{m}_q e^{i\gamma_q} = R_q \mu + A^*$ is the off–diagonal term in the squark mass matrix, $r = m_{\tilde{g}}^2/M_{\tilde{q}}^2$, and $K(r) = -[1 + 5r + 2r(2+r)\ln r/(1-r)]/2(1-r)^3$. Here $m_{\tilde{g}}(M_{\tilde{q}})$ is the gluino (squark) mass, $R_q = \cot\beta\,(\tan\beta)$ for an isospin $T_{3q} = +\frac{1}{2}(-\frac{1}{2})$ quark, and $\tan\beta$ is the usual ratio of the vacuum expectation values of the hypercharge $+1$ and hypercharge $-1$ Higgs scalars. We shall assume throughout a common, complex, dimensionless parameter $A$ for the trilinear terms. Thus, if $|A| \ll |\mu|$ and $\sin\gamma_u \sim \sin\gamma_d$,

$$d_n^G \propto \sum_{q \in n} Q_q R_q m_q (\Delta q)_n . \qquad (2)$$

In Figure 1 we plot, as a function of tan$\beta$, the ratio of the values $d_n^G$ assuming NQM and EMC values of the spin fractions $\Delta q$. Note that the values are of opposite sign. As can be seen there, the ratio can be fairly large for tan$\beta \lesssim 2$ and for the central EMC value of $\Delta s$. Moreover, the ratio can be large for all values of tan$\beta$ if $\Delta s = -0.08$, which is just $1\sigma$ below the central EMC value: $d_n$ is fairly sensitive to the value of $\Delta s$ because $m_s \gg m_d$. In the opposite case that $|A| \gg |\mu|$ (or the off–diagonal terms $\overline{m}_q e^{i\gamma_q}$ are all similar if $A$ is not common), $d_n^G \propto \sum_{q \in n} Q_q m_q (\Delta q)_n$ and the ratio is large as well, $|d_{n,NQM}^G / d_{n,EMC}^G| \approx 4.3$ (28 for $\Delta s = -0.08$).

The chargino–exchange contribution to $d_q$ is given by (again, assuming a common squark mass)

$$d_q^C/e = \frac{\alpha_{EM}}{4\pi \sin^2\theta_W} \left( \frac{M_2 |\mu| \sin\theta_\mu}{m_{\omega_2}^2 - m_{\omega_1}^2} \right) R_q \frac{m_q}{M_{\tilde{q}}^2} \sum_{i=1}^{2} (-1)^i [Q_{\tilde{q}'} I(r_i) + (Q_q - Q_{\tilde{q}'}) J(r_i)] , \qquad (3)$$

where $\theta_\mu$ is the phase of $\mu$, $r_i = m_{\omega_i}^2 / M_{\tilde{q}}^2$, $m_{\omega_2}(m_{\omega_1})$ is the heavy (light) chargino mass, $q'$ refers to the doublet partner of $q$, $I(r) = [1 + r + 2r \ln r/(1-r)]/2(1-r)^2$, and $J(r) = [3 - r + 2r \ln r/(1-r)]/2(1-r)^2$. Since the factor in square brackets is of opposite sign and approximately equal magnitude for quarks of opposite isospin (differing by $\lesssim 10\%$ for $r_1 \leq 1$), in this case we have approximately

$$d_n^C \propto \sum_{q \in n} 2T_{3q} R_q m_q (\Delta q)_n . \qquad (4)$$

In Figure 2 we plot, as a function of tan$\beta$, the ratio of the values $d_n^C$ assuming NQM and EMC values of the spin fractions $\Delta q$. As can be seen there, the results are qualitatively the same as in Fig. 1.

These results, for the simple neutralino sector we have assumed, help illustrate and explain the results of our more general numerical analysis of the MSSM, to which we now turn. There are four neutralinos that contribute to $d_q$ in this model in addition to the charginos and the gluino. We have diagonalized numerically the neutralino mass matrix to find the neutralino composition in the gaugino–Higgsino basis. From this composition, which now depends on the parameters $\mu, M_2, \tan\beta$ and $\theta_\mu$, we determine the contribution of each neutralino to the quark moment $d_q$. We do not assume a common squark mass as before. Rather, we use the squark compositions and masses determined by the diagonalization of the squark mass matrix. In addition, in order to have a more realistic squark mass spectrum, we allow for radiative corrections to the soft supersymmetry breaking masses assuming a common sfermion mass $m_0$ at the GUT scale. Thus, we have differing left and right soft supersymmetry breaking masses. The ratio of $d_n$ calculated with NQM values of the spin fractions $\Delta q$ to $d_n$ calculated with EMC values, $|d_n^{NQM}/d_n^{EMC}|$, depends on many parameters now. In Figure 3 we exemplify our results in the usual $(\mu, M_2)$ plane, for $\tan\beta = 2$ and $\theta_A = 90°$. Here we would expect, on the basis of our previous results, to find a ratio $\gtrsim 1.7$ throughout the plane. We show the ratio $|d_n^{NQM}/d_n^{EMC}|$ for $A = m_0 = 300$ GeV with, in the top (bottom) panel, $\theta_\mu = 20°$ (160°). The light (darker, darkest) shaded areas correspond to regions where $|d_n^{NQM}/d_n^{EMC}| > 1.5$ (3, 6). There are only light–shaded and unshaded areas in the region $\mu < 500$ GeV or $M_2 > 500$ GeV, and this is so throughout the ($\mu < 1000$ GeV, $M_2 < 1000$ GeV) plane for much larger $m_0$.

We find that only in a small fraction of the unshaded regions is the ratio smaller than unity. There is only a very narrow band in the unshaded region of the bottom panel of Fig. 3 (running parallel to the darkest shaded area) where the ratio is substantially smaller than unity. In that band, as well as in the darkest shaded areas of Fig. 3, there is a near cancellation between the chargino and gluino contributions. However, the near cancellation occurs over a much wider area for EMC values of the spin fractions because both contributions are made small, and therefore the difference of small numbers makes for a naturally small $d_n$. For example, in the bottom panel of Fig. 3 $|d_n^{NQM}/d_n^{EMC}| < 0.1$ only in a band of width $\Delta\mu \approx 15$ GeV for $M_2 = 50$ GeV, whereas $|d_n^{NQM}/d_n^{EMC}| > 10$ in a band $\Delta\mu \approx 80$ (125, 750) GeV for $\Delta s = -0.11$ (−0.1, −0.08).

Thus, our previous results are confirmed qualitatively by this more complete treatment of the MSSM. Moreover, we draw the reader's attention to the very dark areas, where even for $\tan\beta = 2$ and $\Delta s = -0.11$ there can be a substantially smaller $d_n$ for relatively light squark masses and large $\theta_A$! This contrasts with the typical situation if one assumes the NQM of the nucleon structure. In that case, the low value of $d_n$ required by experiment ($|d_n| < 1.1 \times 10^{-25}$ $e$ cm [7]) can only arise for relatively heavy (mass $\sim$ TeV) squark masses if the CP–violating phases are $O(1)$ [6]. Such large masses, however, do not allow the neutralino LSP to be a viable dark matter candidate because its relic density would far exceed the observational constraints on the mass density of the universe. Recently [2], however, it has been pointed out that there is some room in the MSSM for a dark matter $\tilde{B}$ even for relatively large CP–violating phases, provided $\tilde{B}$ is sufficiently heavy. In Fig. 3, on the other hand, we see that even for a light neutralino (small $M_2$) there is the possibility of a sufficiently small $d_n$ for relatively light squark masses, and that $d_n$ can be small over a much wider region of parameter space.

One possible source of uncertainty in the value of $d_n$ in the dark regions is that the other contributions might prevent $d_n$ from being small. However, on quite general grounds $d_n$ must be proportional to the neutron spin, and thus the only other quark contribution that might not depend on the spin fractions would depend on orbital angular momentum, which appears to be small [8].

We conclude on the basis of the above arguments that a more realistic model of the quark structure of the nucleon spin indicates that $d_n$ is smaller than previously thought in supersymmetric theories. This weakens constraints on CP-violating parameters in supersymmetric models, and further increases the upper limit on the LSP mass inferred from its relic density by calculations that allow for CP–violating phases. We have also shown that $d_n$ could be substantially smaller if the LSP is a relatively light $\tilde{B}$, thus raising the possibility that a *light* neutralino could constitute the cold dark matter without requiring particularly small CP–violating phases in the theory.

The work of RF is supported by an NSF grant. He thanks the hospitality of the Theory Division at CERN while this work was started.

# FIGURE CAPTIONS

**Figure 1** – The ratio $-d^G_{n,NQM}/d^G_{n,EMC}$, obtained using Eq. (2), of the gluino contribution to $d_n$ in the simple model with a $\tilde{B}$ LSP calculated with NQM values of the spin fractions $\Delta q$ ($\Delta u = \frac{4}{3}$, $\Delta d = -\frac{1}{3}$, $\Delta s = 0$) to the contribution calculated with EMC values ($\Delta u = 0.82$, $\Delta d = -0.44$, and $\Delta s = -0.11$ ($-0.08$) for the solid (dotted) curve).

**Figure 2** – The ratio $-d^C_{n,NQM}/d^C_{n,EMC}$, obtained using Eq. (4), of the chargino contribution to $d_n$ in the simple model with a $\tilde{B}$ LSP calculated with NQM values of the spin fractions $\Delta q$ to the contribution calculated with EMC values. The values of the spin fractions are as in Figure 1.

**Figure 3** – The ratio $|d^{NQM}_n/d^{EMC}_n|$ of $d_n$ in the MSSM calculated with NQM values of the spin fractions $\Delta q$ to $d_n$ calculated with EMC values. The values of the spin fractions are as in Figure 1. We have assumed $\theta_\mu = 20°$ ($160°$) in the top (bottom) panel. In the light (dark, darker) shaded areas, $|d^{NQM}_n/d^{EMC}_n| > 1.5$ (3, 6).

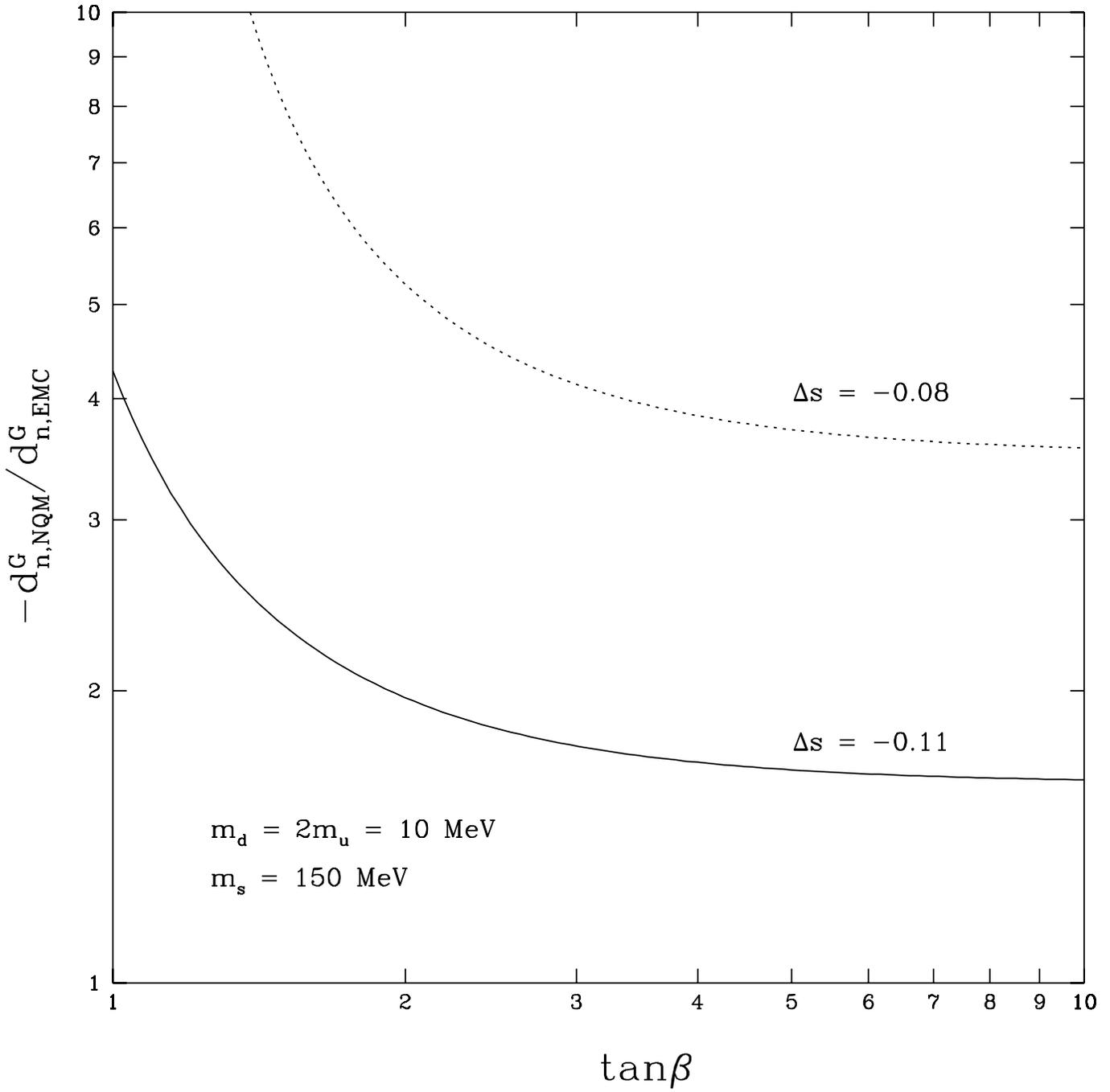

FIGURE 1

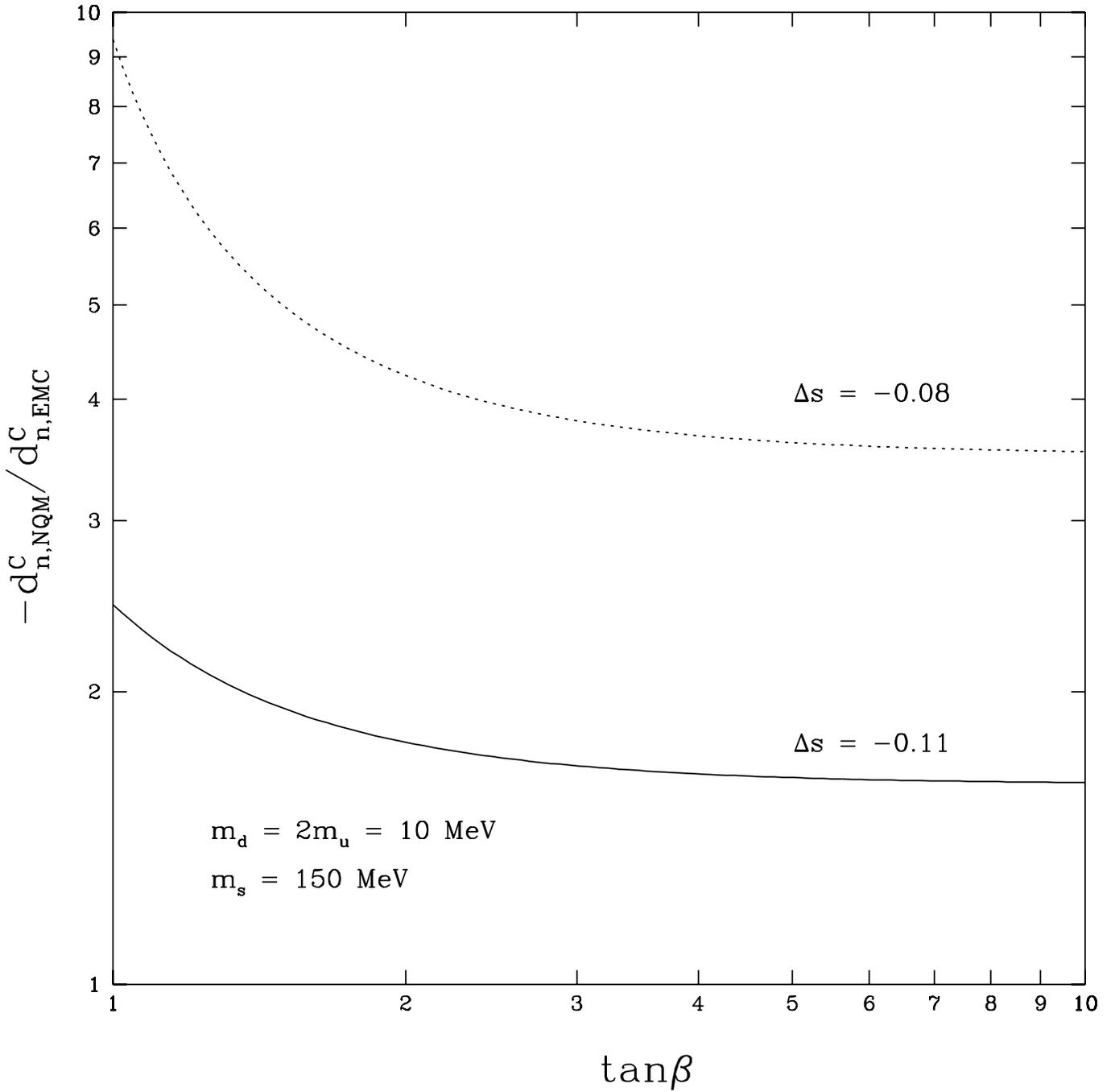

FIGURE 2

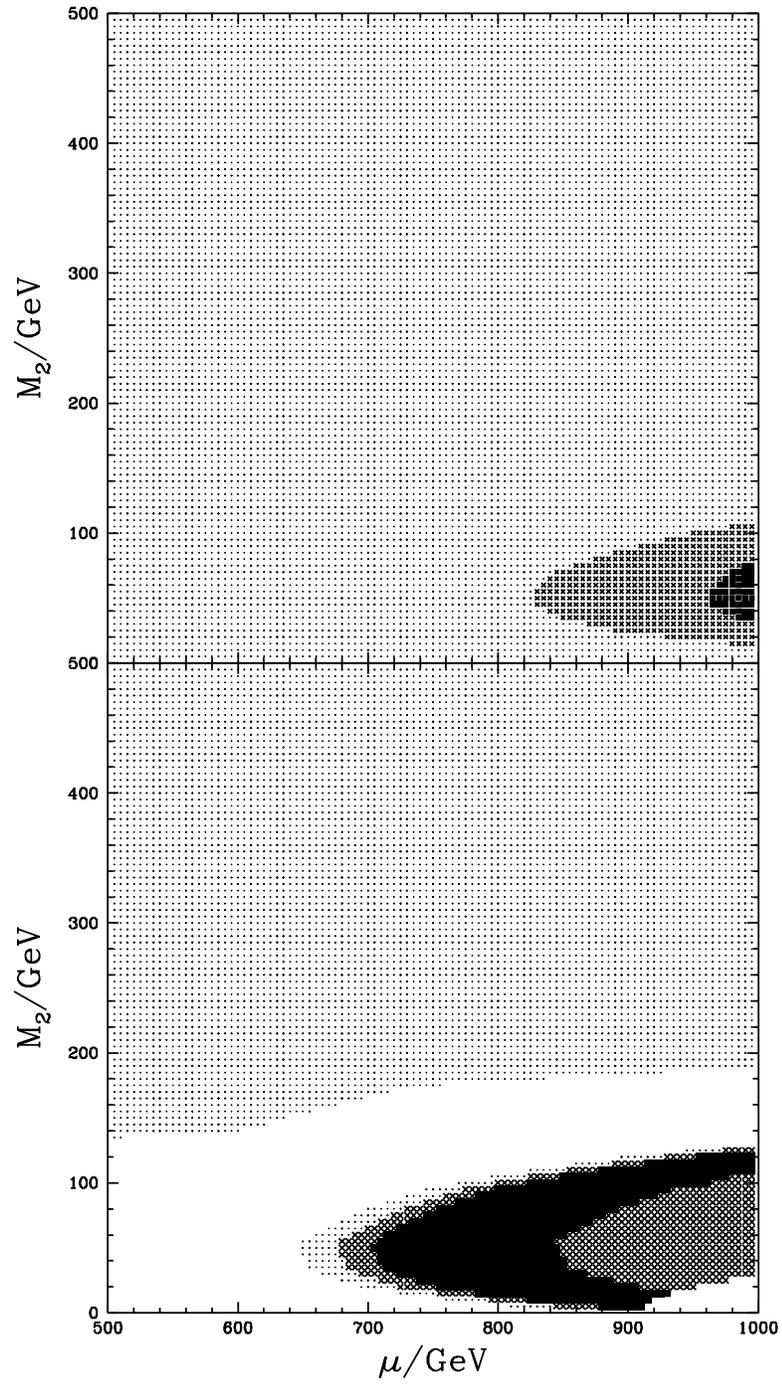

FIGURE 3